\pdfoutput=1 
\documentclass[conference,letterpaper]{IEEEtran}

\IEEEoverridecommandlockouts
\IEEEpubid{978-1-6654-5223-6/23\$31.00~\copyright~2023 IEEE}

\usepackage{orcidlink}
\hypersetup{pdfborder={0 0 0}}
 
\usepackage[utf8]{inputenc}
\usepackage{cite}
\usepackage{amsmath,amssymb,amsfonts}
\usepackage{algorithmic}
\usepackage{graphicx}
\usepackage{textcomp}
\usepackage{xcolor}
\usepackage{tabularx} 

 \usepackage{subcaption} 
 \usepackage{url}  
 \usepackage{booktabs,multirow}
\graphicspath{{images/}} 

\usepackage{xcolor}
\usepackage{mdframed}

\usepackage{flushend}

\def\BibTeX{{\rm B\kern-.05em{\sc i\kern-.025em b}\kern-.08em
    T\kern-.1667em\lower.7ex\hbox{E}\kern-.125emX}}
\begin{document}

\newcommand{\GH}{GitHub}
\newcommand{\SL}{Simulink}
\newcommand{\MW}{MathWorks}

\newcommand{\etal}{et al.}

\newcommand{\xTable}[1]{Table~\ref{#1}}
\newcommand{\xFigure}[1]{Figure~\ref{#1}}

\newcommand{\CorpusCurated}{SC}
\newcommand{\corpusSLNET}{SLNET}
\newcommand{\CorpusBoll}{\CorpusCurated\textsubscript{20}}  

\newcommand{\CorpusCuratedRep}{\CorpusCurated\textsubscript{R}}
\newcommand{\CorpusBollRep}{\CorpusBoll\textsubscript{R}}
\newcommand{\CorpusBollEvol}{\CorpusBollRep\textsubscript{Evol}}
\newcommand{\CorpusSLNETEvol}{\corpusSLNET\textsubscript{Evol}}

\newcommand{\CC}[1]{\emph{\textbf{{\color{red} [CC says: #1]}}}}
\newcommand{\SLS}[1]{\emph{\textbf{\color{magenta}[SS says: #1]}}}
\newcommand{\SAC}[1]{\emph{\textbf{\color{orange}[SAC says: #1]}}}

\mdfdefinestyle{MyFrame}{%
    linecolor=black,
    outerlinewidth=2pt,
    innertopmargin=4pt,
    innerbottommargin=4pt,
    innerrightmargin=4pt,
    innerleftmargin=4pt,
    leftmargin = 4pt,
    rightmargin = 4pt,
    backgroundcolor=gray!10!white
}

\newcounter{insight}
\newenvironment{insight}{
    \stepcounter{insight}
    \begin{mdframed}[style=MyFrame]\itshape
    \textbf{Finding~\theinsight:}
    }
    {
    \end{mdframed}
    }

\newcommand{\totalGithubproject}{1,284}
\newcommand{\totalGithubprojectwithLicense}{232}
\newcommand{\totalGithubprojectinSLNetLegalwithduplicatesanddummy}{231} 
\newcommand{\totalGithubprojectinSLNetLegal}{225}
\newcommand{\totalmodelFilesinGithub}{6,344}
\newcommand{\totalmodelFilesinGithubwithlicense}{2,132} 
\newcommand{\totalmodelFilesinGithubinSLNetLegal}{2,088}  
\newcommand{\totalcompilablemodelGithub}{5,796}
\newcommand{\totalcompilablemodelGithubLegal}{2,048}
\newcommand{\blockCountGithubAll}{2,908,797}
\newcommand{\blockCountGithubLegal}{600,342}

\newcommand{\totalMATCproject}{2,941}
\newcommand{\totalMATCprojectwithLicense}{2,746}
\newcommand{\totalMATCprojectinSLNetLegalwithduplicatesanddummy}{2,728}
\newcommand{\totalMATCprojectinSLNetLegal}{2,612}
\newcommand{\totalmodelFilesinMATC}{11,588}
\newcommand{\totalmodelFilesinMATCwithlicense}{10,039}
\newcommand{\totalmodelFilesinMATCSLNetLegal}{7,029}
\newcommand{\totalcompilablemodelMATC}{8,393}
\newcommand{\totalcompilablemodelMATCLegal}{7,122}
\newcommand{\blockCountMatCAll}{4,764,383}
\newcommand{\blockCountMATCLegal}{4,407,700}

\newcommand{\SLNETallTotalProject}{4,225}
\newcommand{\SLNETlicenseTotalProject}{2,978}
\newcommand{\SLNETlegalTotalProjectswithduplicatesanddummy}{2,959}
\newcommand{\SLNETlegalTotalProject}{2,837}

\newcommand{\SLNETallTotalModel}{17,932}

\newcommand{\SLNETlegalTotalModel}{9,117}
\newcommand{\SLNETalltotalcompilablemodel}{14,189}
\newcommand{\SLnetLegaltotalcompilablemodel}{9,170}
\newcommand{\SLNETnoMetricModels}{88}

\newcommand{\SLNETalltotalBlock}{7,673,180}
\newcommand{\SLnetLegaltotalBlock}{5,008,042}

\newcommand{\modeloverthousand}{983}
\newcommand{\modelovertenthousand}{60}
\newcommand{\modeloverthousandblocksusingnonimportedcount}{149}
\newcommand{\SlNetprojectCount}{2,949}

\title{Replicability Study: Corpora For Understanding Simulink Models \& Projects}

\author{\IEEEauthorblockN{Sohil Lal Shrestha \orcidlink{0000-0002-0837-8388}}
\IEEEauthorblockA{\textit{Computer Science \& Eng. Dept.} \\
\textit{University of Texas at Arlington}\\
Arlington, TX 76019, USA \\
sohil.shrestha@mavs.uta.edu}
\and
\IEEEauthorblockN{Shafiul Azam Chowdhury \orcidlink{0000-0001-9019-6067}}
\IEEEauthorblockA{\textit{Computer Science \& Eng. Dept.} \\
\textit{University of Texas at Arlington}\\
Arlington, TX 76019, USA \\
shafiulazam.chowdhury@mavs.uta.edu}
\and
\IEEEauthorblockN{Christoph Csallner \orcidlink{0000-0003-0896-6902}}
\IEEEauthorblockA{\textit{Computer Science \& Eng. Dept.} \\
\textit{University of Texas at Arlington}\\
Arlington, TX 76019, USA \\
csallner@uta.edu}
}

\maketitle

\begin{abstract}

\emph{Background:} Empirical studies on widely used model-based development tools such as MATLAB/\SL{} are limited despite the tools' importance in various industries.

\emph{Aims:} 
The aim of this paper is to investigate the reproducibility of previous empirical studies that used \SL{} model corpora and to evaluate the generalizability of their results to a newer and larger corpus, including a comparison with proprietary models.   

\emph{Method:} 
The study reviews methodologies and data sources employed in prior \SL{} model studies and replicates the previous analysis using \corpusSLNET{}. In addition, we propose a heuristic for determining code-generating \SL{} models and assess the open-source models' similarity to proprietary models. 

\emph{Results:} 
Our analysis of \corpusSLNET{} confirms and contradicts earlier findings and highlights its potential as a valuable resource for model-based development research. We found that open-source \SL{} models follow good modeling practices and contain models comparable in size and properties to proprietary models. We also collected and distribute 208~git repositories with over 9k~commits, facilitating studies on model evolution.

\emph{Conclusions:} 
The replication study offers actionable insights and lessons learned from the reproduction process, including valuable information on the generalizability of research findings based on earlier open-source corpora to the newer and larger \corpusSLNET{} corpus. The study sheds light on noteworthy attributes of \corpusSLNET{}, which is self-contained and redistributable. 

\end{abstract}

\begin{IEEEkeywords}
reproducibility, replication, \SL{}, open science, code generation, \SL{} model
\end{IEEEkeywords}

\section{Introduction}


There are only a few empirical studies of open-source MATLAB/\SL{} artifacts, maybe due to a widespread perception that open-source \SL{} artifacts are typically small, do not represent closed-source development, and are often hard to acquire~\cite{simulink:noopensourceModels:Matinnejad,simulink:noopensourceModels:mutation:Rao,simulink:slicing:Jiang,Chowdhury20SLEMI,chowdhury2018icse}. Most empirical \SL{} studies to date have instead relied on academic-industry collaborations---to get access to large closed-source \SL{} artifacts~\cite{BertramMRRW17ComponentAndConnector}. Most empirical results on \SL{} development and artifacts are thus based on case-studies of closed-source artifacts that (even when providing detailed experimental design descriptions and measurement tools) are hard to reproduce or replicate~\cite{simulink:onreplicability:Boll}.


It is well-known how important replication is for scientific progress. Successful experiments need to be cross-validated under different conditions before they can be considered a part of science and interpreted with confidence~\cite{campbell_stanley}. Working towards large open-source \SL{} corpora and empirical results that are easier to reproduce and replicate are thus important goals, given how widely \SL{} is used in industry in safety-critical domains such as automotive and healthcare.


Towards these goals, recent initial work created via manual mining a first large corpus (which we call \CorpusCurated{}~\cite{Chowdhury18Curated}) of open-source \SL{} models and investigated modeling practices on a re-collected version of that corpus (\CorpusBoll{}~\cite{simulink:corpus:boll2021characteristics}). The work found that some of these manually-collected \SL{} models are suitable for empirical research, based on model metrics analysis and a qualitative assessments by a domain expert~\cite{simulink:corpus:boll2021characteristics}. Follow-up work automated \SL{} model collection, yielding the larger \corpusSLNET{} corpus that also allows redistribution~\cite{paper:slnet}. However we are not aware of earlier work that either characterizes this larger \corpusSLNET{} corpus or uses it to replicate earlier empirical studies of \SL{} models.


\IEEEpubidadjcol

We thus first reproduce studies that are based on the initial \CorpusCurated{} large-scale \SL{} model corpus, identifying inconsistencies in the original studies. We then replicate results of the earlier studies using the newer and larger \corpusSLNET{} corpus. By re-running the original study designs, we found inconsistencies between the experimental results and the ones reported in the paper, attributable to oversight and incomplete documentation. Our replication study using \corpusSLNET{} confirmed several previous findings, such as the low utilization of model references and algebraic loops. In contrast to prior work, we only found a weak correlation between cyclomatic complexity and other model metrics. To summarize, this paper makes the following major contributions.
\begin{itemize} 
\item Through empirical data, we identify inconsistencies in earlier empirical \SL{} studies.
\item We characterize the \corpusSLNET{} corpus in relation to earlier corpora of open-source \SL{} models.
\item On \corpusSLNET{} we replicate previous studies, which both confirms and contradicts earlier findings.
\item We collect and distribute 208 SLNET git repositories, containing 9k+ commits including 5k model versions, as artifacts that can be analyzed by the community~\cite{dataset:simulinkanalysisdata}. 
\item Our analysis tool~\cite{doc:slreplicationTool} as well as reproduction and replication data~\cite{dataset:simulinkanalysisdata} are open-sourced and available.      
\end{itemize}

\section{Background}
\label{sec:background}

Using \SL{}'s graphical modeling environment, engineers can design a complex system model as a hierarchical \emph{block diagram}~\cite{doc:simulink}. Each block represents a dynamic system that may take input through its \emph{input ports} and produce output via its \emph{output ports}, either continuously or at specific points in time. A block can be from a \SL{} built-in library~\cite{doc:blocklib}, from a separate \emph{toolbox} library, or a custom \emph{S-function} block defined via ``native'' code (e.g., in C). Blocks pass data to each other via directed \emph{connections} (aka lines). \SL{} is a commercial de-facto standard tool-chain in several domains such as aerospace, automotive, healthcare, and industrial automation.

\begin{figure}[h]
\centering
 \includegraphics[width=0.9\linewidth]{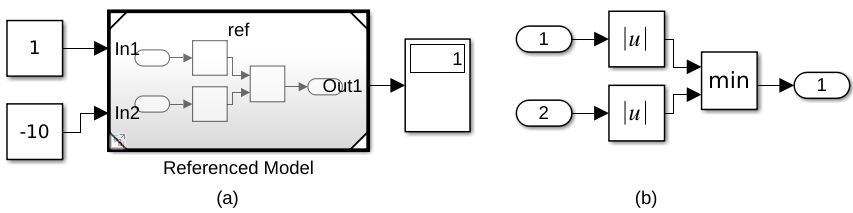}
  \caption{(a) A tiny \SL{} example model, 
  (b) shows the contents of (a)'s referenced model.
  }
  \label{fig:toySimulinkModel}
  \centering
\end{figure} 

\SL{} offers several hierarchy mechanisms, ranging from a \emph{subsystem} block grouping that can only be used in one context to a \emph{model reference} (which essentially calls an independent model via its own well-defined interface and can thus be widely reused)~\cite{doc:SL-HierarchyOptions}. These constructs allow further recursive decomposition, enabling deeply nested models. \xFigure{fig:toySimulinkModel}(a) shows a tiny example hierarchical model that contains a model reference to the \xFigure{fig:toySimulinkModel}(b) referenced model.
Alternatively, the user can use library-linked blocks~\cite{doc:librarylinkedBlocks}, that are references to blocks defined in a custom library~\cite{doc:customLibraries}, that enables reusability and centralized maintenance of block functionality across multiple models.

A \emph{compiled} model can be simulated, where \SL{} successively computes the output of each block over a specified time range using pre-configured numerical \textit{fixed-step} and \textit{variable-step} solvers. In an \emph{algebraic loop}, a block's output can reach its input port in the same simulation step (i.e., without passing through a delay block), which complicates simulation. Besides \textit{normal} mode, \SL{} offers various \textit{accelerator} modes to speed up simulation~\cite{doc:simulation}. With additional toolboxes~\cite{doc:mathworksproducts}, from the model the user can then generate and deploy low-level code to the target hardware.

\subsection{\SL{} Modeling Guidelines and Best Practices}

The MathWorks Advisory Board (MAB) is a group of commercial MathWorks customers that (starting with Daimler, Ford, and Toyota in 2001) publishes guidelines and best practices on developing and maintaining \SL{} models. Besides standardization, these guidelines address key software engineering challenges such as creating models that are well-defined, readable, easy to integrate, and reusable.  

In their current 2020 version~\cite{doc:MABmodelingGuidelines} these guidelines include to
(1)~avoid algebraic loops as they are hard to simulate and cannot be compiled to target hardware,
(2)~use S-functions to implement custom algorithms,
(3)~use subsystems to modularize the model by functional decomposition, and
(4)~use model references to create hierarchies of reusable components.

\subsection{Cyclomatic Complexity \& Size Metrics in \SL{}}

McCabe introduced cyclomatic complexity and argued it corresponds to our intuitive notion of complexity. McCabe also reported on a set of 24~Fortran subroutines with ``high'' ($>$10) cyclomatic complexity. The subroutines' ranking by cyclomatic complexity closely correlated with their ranking by reliability~\cite{cyclodef:mccabe}. With some 9k citations this article has been highly influential in academic software engineering. 

Some five decades later the question of measuring program complexity and program understanding remains an active research area with several recent advances~\cite{Levy21Understanding,Feitelson22Considerations,Chowdhury22Revisiting}. Researchers keep returning to cyclomatic complexity with recent tweaks~\cite{Campbell18Cognitive,Baron20Empirical} and more fine-grained measures~\cite{Ajami19Syntax}. An example controversy was if cyclomatic complexity is just a proxy for program size (e.g., lines of code in a Java- or C-like language)~\cite{Jay09Cyclomatic}, with recent empirical data showing cyclomatic complexity to remain independently valuable~\cite{Landman16CC-SLOC}.

For \SL{}, recent work has shown the value of size metrics (i.e., block count), e.g, metric outliers yield interesting findings~\cite{Schroeder16Unveiling}. Such results are also eventually reflected in industry practices. For example, while the MAB industry board's 2001 \SL{} guidelines did not yet mention size metrics, the current 2020 version contains a recommendation ($\leq$60 LOC / function)~\cite{doc:MABmodelingGuidelines}. However neither MAB guideline version mentions McCabe or cyclomatic complexity yet.

For calculating a metric, \SL{} basically first flattens a given model into a single hierarchy level, essentially ``inlining'' both subsystems and referenced models. So if two blocks in a model refer to the same referenced model, for metric calculations the referenced model will appear in the flattened model twice. \SL{} has an option to also similarly (recursively) inline the contents of (any) library blocks and prior work is split on activating this option when reporting metric results.

While a block diagram does not represent a procedural language's control-flow graph, \SL{} still has several block types that provide control-like functionality. For example, the value a multiport-switch block receives on its first input port selects which of the remaining input ports the block will forward to its output port~\cite{doc:multiportswitch} (which corresponds to a procedural switch or nested if construct). \SL{} thus first defines the cyclomatic complexity of each built-in block as the number of the block's conceptual branching decisions (i.e., mostly zero or one) and then sums up the cyclomatic complexity of all blocks in a given (flattened) model~\cite{doc:SL-McCabe}.

\subsection{Scope of Empirical Studies of \SL{} Models}
 \label{original_info}

The limited availability of repositories with large numbers of freely accessible \SL{} models has restricted empirical studies that seek to understand \SL{} model characteristics and metrics~\cite{Balasubramaniam:2020,simulink:hu2012quality,simulink:Marta}. For example, 
Dajsuren \etal{}~\cite{simulink:akasoftware_Dajsuren} investigated model metrics including cohesion and coupling using small subset of \SL{} models. 

Open-source \SL{} models are generally considered insufficient to meet the high industry standards required for meaningful results~\cite{simulink:onreplicability:Boll}. To address this issue, Altinger \etal{}~\cite{Altinger15Novel} published metrics from three proprietary  \SL{} models for researchers to analyze. However, the dataset is no longer available.  Schroeder \etal{} studied 65 proprietary automotive \SL{} models and found via interviews that engineers preferred simple size metrics such as block count over structural metrics to capture model complexity~\cite{Schroeder15Comparing}.

\subsection{\CorpusCurated{}: First Corpus of Open-Source \SL{} Models}

Via a two-stage process Chowdhury \etal{} created what we call \CorpusCurated{}, the first corpus of freely available \SL{} models~\cite{chowdhury2018icse,Chowdhury18Curated}. First~\cite{chowdhury2018icse}, the research team collected 391 models, i.e., 41 of the MathWorks's tutorial models the team considered to not be ``toy'' examples, the open-source models from MATLAB Central that were most popular (by ratings or downloads), GitHub keyword search results, and 28 models from academic papers, colleagues, and Google searches. Second, the team added the \SL{} models of 12 SourceForge repositories and of the 96 most-downloaded MATLAB Central projects, yielding a study of a total of 1,071 \SL{} models~\cite{Chowdhury18Curated}.

\CorpusCurated{} classifies its 1,071 models as tutorial (41),  simple (442), advanced (452), and other (136). The distinction between simple and advanced is determined heuristically: any \GH{} project with forks or stars and any MATLAB Central project that are not academic assignment are labeled ``Advanced''. Models shipped with MATLAB/\SL{} are labeled ``Tutorial'', while models from other sources are labeled 'Other'.

Overall, \CorpusCurated{} collects \SL{} models of projects that (at least partially) are selected and labeled manually. While initially ``only'' providing project URLs~\cite{chowdhury2018icse}, the full corpus~\cite{Chowdhury18Curated} includes \SL{} model files, metadata, and collection tools and is stored on a Google Drive directory linked from the project's GitHub homepage.

Analyzing the corpus with \SL{} R2017a, the work found good modeling practices such as model referencing were not widely used. The work found MathWorks's cyclomatic complexity to be at most moderately correlated\footnote{The earlier work discussed in this paragraph and our own analysis all use Kendall's $\tau$ at a 0.05 significance level and follow a recent labeling of subsequent $|\tau|$ ranges at that level, i.e.: ``weak'' below 0.4, then ``moderate'' to below 0.7, ``strong'' to below 0.9, etc.~\cite{kendaltaurange:esem:MamunBH19}} with various other model metrics. The correlation was strongest (0.55) for the model's maximum hierarchy depth, followed by the model's number of contained subsystems (NCS). This contrasted with an earlier study by Olszewska~\etal{}~\cite{simulink:Marta}, which showed strong (0.73) correlation between MathWorks's cyclomatic complexity and the model's number of contained subsystem (NCS).

\subsection{\CorpusBoll{}: \CorpusCurated{} Projects Recollected in 2020}

In August 2020---some three years after \CorpusCurated{} was published~\cite{Chowdhury18Curated}, Boll \etal{} (a research team acting independently of Chowdhury \etal{}) collected what we call \CorpusBoll{}~\cite{simulink:corpus:boll2021characteristics}, i.e., the latest \SL{} model versions of \CorpusCurated{}'s \SL{} projects, yielding 1,734 \SL{} models. \SL{} models, metadata, and the team's collection tools are preserved on Figshare~\cite{dataset:boll}.

The work evaluated \CorpusBoll{}'s suitability for empirical model-based research, analyzing each \CorpusBoll{} project's domain, origin, and model metrics. The work also proposed a heuristic for identifying models configured for code generation.
The paper's analysis found that the majority of \CorpusBoll{} models were inadequate for most empirical research, but identified a few mature models.
The work also noted that some \CorpusBoll{} \GH{} projects' characteristics (e.g., a high number of commits and collaborators) suggest potential for evolution research.

\subsection{\corpusSLNET{}: Largest Known \SL{} Corpus}

In February 2020 Shrestha \etal{} collected the \corpusSLNET{} corpus~\cite{paper:slnet}, which addresses key issues of \CorpusCurated{} and \CorpusBoll{} (i.e., manual project selection and unclear project licenses), yielding the first redistributable corpus of open-source \SL{} models. Specifically, \corpusSLNET{} collects \SL{} projects from the GitHub API and from MATLAB Central's RSS feed and does not include projects without \SL{} model files, known model generators and their synthetic models, projects that do not have an appropriate license, potentially duplicate projects (via bijection of the projects' models' metrics), and projects whose models all have zero blocks, yielding 9,117 \SL{} models. \SL{} models, metadata, and the team's collection tools are preserved on Zenodo~\cite{dataset:slnet,tool:SlnetMetrics,tool:Slnet-miner}.

Combining models from the two largest collections of open-source \SL{} models (\GH{} and MATLAB Central), \corpusSLNET{} is 8~times larger than the largest previous corpus of \SL{} models (\CorpusCurated{}). In March 2023 we confirmed that other hosting sites (still) contain significantly fewer public \SL{} repositories (i.e., we could only find 52~\SL{} projects on SourceForge and one on GitLab).

\section{Research Design}

Our goal is to gain a deeper understanding of the reproducibility and replicability in model-based development research, particularly regarding Simulink models, as emphasized in a recent literature review~\cite{simulink:onreplicability:Boll}. The literature review identified a single study that conducted a large-scale empirical investigation, emphasizing open science, i.e., \CorpusCurated{}~\cite{Chowdhury18Curated}. Subsequently, members of the literature review team undertook their own investigation, by collecting the latest version of the models of the same corpus, i.e., \CorpusBoll{}~\cite{simulink:corpus:boll2021characteristics}.

The recently released \corpusSLNET{} corpus~\cite{paper:slnet} has rectified limitations of the two existing corpora, allowing us to replicate the results of earlier empirical studies. Thus, we perform a sample study utilizing the existing corpora and employ a statistical learning strategy to generalize the findings of prior studies on a smaller dataset to a larger dataset~\cite{sixGeneralizationStrategy,StolF18ABCofSEResearch}. As such, our replication efforts serve a confirmatory purpose.

To structure our study effectively, we have formulated two primary research questions that center around reproducibility and replication. 
\begin{itemize}
    \item[I] What challenges and implications arise when attempting to reproduce model-based development research, specifically for Simulink models?
    \item[II]To what extent can we generalize prior studies' findings to a dataset that is open-source or larger?
    \begin{itemize}
\item[RQ1] In terms of basic \SL{} model metrics, how does \corpusSLNET{} compare with earlier open-source corpora and what we know about industrial models?
\item[RQ2] Is \corpusSLNET{} suitable for empirical studies of \SL{} projects and their change histories?
\item[RQ3] How do empirical results obtained on smaller open-source corpora and closed-source industry models carry over to the larger \corpusSLNET{} corpus?
\end{itemize}
    \end{itemize}

\begin{figure}[h!t]
\centering
\includegraphics[trim = .2in .2in .2in .2in, clip, width=\linewidth]{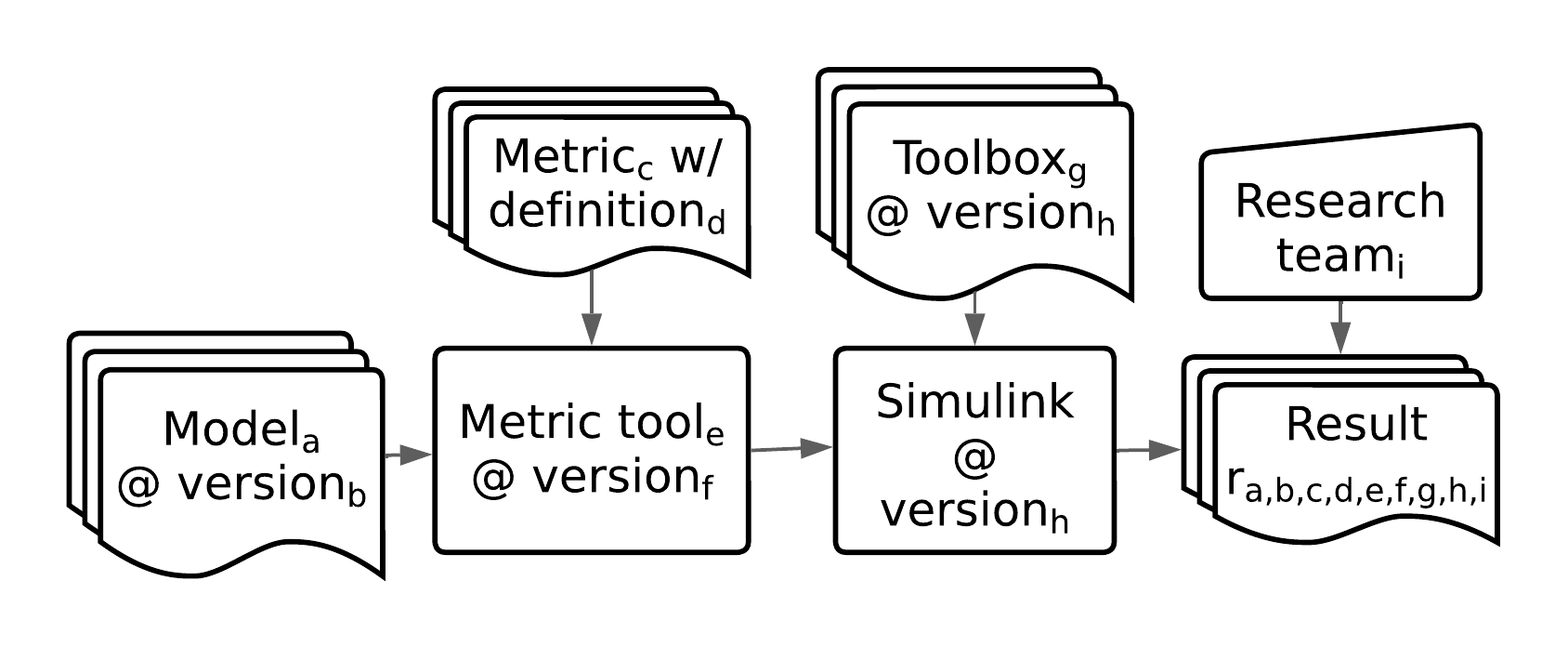} 
  \caption{
  Parameters \emph{a} through \emph{i} for reproducing and replicating results on \SL{} models. Relative to earlier studies (and unless noted otherwise), for reproduction we only varied \emph{i} and for replication we only varied \emph{a,b,i}.
  }
  \label{fig:experiment-params}
  \centering
\end{figure} 

 Figure~\ref{fig:experiment-params} applies ACM's guidelines on reproducibility (``different team, same experimental setup'') and replicability (``different team, different experimental setup'') to empirical studies of \SL{} models and summarizes the relevant variables. The following sections point out where we had to deviate from this model (e.g., when an exact earlier corpus is no longer available for exact reproduction).

\section{Corpora to Reproduce \& Replicate Results}
\label{sec:orignalstudychange}

\begin{table}[h]
\centering
\caption{Overview of three existing (top) plus our four new or re-collected corpora (bottom) of open-source \SL{} models; cut-off~=~date of latest model version in corpus; $\times$~=~cannot distribute due to unclear licenses.
}
\begin{tabular}{llrc} 
\toprule
Corpus & Version of \SL{} Models & Cut-off & Data \\
\midrule
\CorpusCurated{}    & Original corpus   & 2017 & \cite{doc:corpussimdoc}\\
\corpusSLNET{}      & Larger corpus     & Feb '20 & \cite{dataset:slnet} \\
\CorpusBoll{}       & \CorpusCurated{} re-collected at later version & Aug '20 & \cite{dataset:boll} \\
\midrule
\CorpusCuratedRep{} & \CorpusCurated{} re-collected at \CorpusCurated{}'s version  & 2017 & $\times$\\ 
\CorpusBollRep{}    & \CorpusBoll{} completed at \CorpusBoll{}'s version & Aug '20 & $\times$\\
\CorpusBollEvol{} & \CorpusBollRep{} GitHub projects' Git histories & Apr '23 & $\times$\\
\CorpusSLNETEvol{} & \corpusSLNET{} GitHub projects' Git histories & Apr '23  & \cite{dataset:simulinkanalysisdata}\\
\bottomrule
\end{tabular}
\label{table:corpora}
\end{table}

Table~\ref{table:corpora} summarizes the corpora of this study. Boll~\etal{}~\cite{simulink:corpus:boll2021characteristics} highlighted that the \CorpusCurated{} study results had several inconsistencies and Shrestha~\etal{}~\cite{paper:slnet} claimed earlier corpora suffer from unintended human errors and bias. Since both claims lacked sufficient empirical evidence, we attempted to reproduce these studies.

\subsection{\CorpusCuratedRep{} Corpus to Reproduce \CorpusCurated{} Results}
\label{reproduce:SLC}

To reproduce the \CorpusCurated{} study results, we downloaded all models and metadata from \CorpusCurated{}'s Google Drive~\cite{doc:corpussimdoc}, yielding 1,347 models. This did not include all of the original study's 1,071 models, as the \CorpusCurated{} distribution excludes 169 models for their unclear licenses. We use \CorpusCurated{}'s source metadata (for 862 of 1,071 models \CorpusCurated{} lists project URL and version, models studied within the project, and MATLAB version requirements) and retrieve 142 of 169 of these unclearly-licensed models from GitHub (at the same version as in \CorpusCurated{}). 

For 40 of 1,071 models the download included multiple model versions but the metadata did not specify which version was used in the \CorpusCurated{} study. Since \CorpusCurated{} only provides aggregated model metrics (instead of per-model measurements), we could not disambiguate same-name models via comparing the metrics. After receiving confirmation from the \CorpusCurated{} team, we add all 113 potential model name matches from the \CorpusCurated{} download, yielding 1,117 models in \CorpusCuratedRep{} (but still missing 27 of 1,071 now inaccessible GitHub models).

Due to the above model name ambiguity (or human error in \CorpusCurated{} creation), 5 of 1,071 models are now categorized as both Simple and Advanced. Since the \CorpusCurated{} study reported results per model category, we focused our reproduction on the one category not affected by the above missing/duplicate model issues, i.e., the 41 models labeled ``tutorial''. Since these 41 models ship with \SL{} and we have access to earlier \SL{} releases, it was straight-forward to reproduce the \CorpusCurated{} study results on the same version of the same models on the same \SL{} version as the \CorpusCurated{} study.

The \CorpusCurated{} paper states that some reported metric results come from a third-party tool~\cite{tool:howmanyblocks}. But we found the tutorial models' reported metrics instead exactly match the results of only running the \CorpusCurated{} metrics tool (which calls the Simulink API~\cite{doc:simulinkcheck}). Specifically, we ran the SLNET-Metrics tool~\cite{tool:SlnetMetrics} as it can run \CorpusCurated{}'s metric tool in the \SL{} toolbox configuration~\cite{doc:artifactsdocslforge} the \CorpusCurated{} study used, yielding the reported 10,926 blocks (as opposed to 10,391 the other tool returns~\cite{tool:howmanyblocks}). After this calibration on the tutorial models we ran the \CorpusCurated{} tool in the same configuration on the rest of \CorpusCuratedRep{}.

Finally, we clarified with the \CorpusCurated{} team \CorpusCurated{}'s ``S-function reuse rate'', which \CorpusCurated{} defined to approximate how often a model contains an S-function it contains elsewhere. The metric basically counts how many S-function blocks in a model have the same name.
For example, if a model contains four S-function blocks, three named ``a'' and one named ``b'', the reuse rate would be (3-1 + 1-1)/(3+1) = 0.5. \CorpusCurated{} reported a median reuse rate below 0.5\%. Our result on \CorpusCuratedRep{} being much higher triggered an interaction, in which the \CorpusCurated{} team confirmed that the \CorpusCurated{} paper mistakenly added the percentage symbol.

\subsection{\CorpusBollRep{} \& \CorpusBollEvol{} Corpora to Reproduce \CorpusBoll{} Results}

We obtained the \CorpusBoll{} replication package (v2) from  Figshare~\cite{dataset:boll}, which contains 1,736 models grouped into 194 projects with non-model files removed. The \CorpusBoll{} team categorized projects into four groups based on affiliation: 112 academic, 34 industry-mathworks, 25 industry, and 23 no-information. We included one project with an unknown category in the `no-information' category, yielding \CorpusBollRep{}. 

To extract model metrics, \CorpusCurated{} and \CorpusBoll{} mostly use the \SL{} API, but there are differences. For instance, \CorpusBoll{} counts blocks via sldiagnostics~\cite{doc:sldiagnostic} while \CorpusCurated{} uses Simulink Check~\cite{doc:simulinkcheck} (the counts can differ). Additionally, \CorpusCurated{} uses the \SL{} API for cyclomatic complexity, while \CorpusBoll{} implements McCabe's definition (independent paths).
From the \CorpusBoll{} paper~\cite{simulink:corpus:boll2021characteristics} and our correspondence with the \CorpusBoll{} team we could not reconstruct how \CorpusBoll{} computed project-level cyclomatic complexity.

The remaining model metrics we reproduced using the provided tool and documentation. To run the tool we had to install \SL{} R2020a and the Check toolbox. We observed discrepancies in the results of 11/1736 models, which we attribute to a lack of documentation regarding the exact \SL{} configuration (i.e., toolbox, library, etc).

 The \CorpusBoll{} team analyzed 35~\GH{} projects, but didn't include the necessary git repositories or commit extraction tool in the replication package. We independently developed the tool, and after contacting the authors, they updated their package, but the repositories remained missing. In April 2023 via metadata we obtained 32/35 repositories (``\CorpusBollEvol{}''). 3/35 repositories were no longer online.

\begin{insight}
The \CorpusCurated{} and \CorpusBoll{} replication packages are insufficient to reproduce the original studies' results. \\
    \textbf{Implication:} 
    Authors should host the replication package in permanent archival repositories for long-term access and preservation with documentation, such as Simulink configuration and instruction~\cite{Mendez20OpenScienceInSE}.
\end{insight}

\subsection{\corpusSLNET{} Results \& \CorpusSLNETEvol{} Corpus}

While the \corpusSLNET{} paper does not present any specific study or analysis, it does offer a valuable resource in the form of a corpus of \SL{} models along with associated metadata on their metrics. In our attempt to reproduce \corpusSLNET{}'s metrics, we first downloaded their corpus from Zenodo~\cite{dataset:slnet}, which consists of 225 \GH{} and 2,612 MATLAB Central projects, as well as a SQLite database of metadata.  Following their documentation on \SL{} configuration~\cite{SLNET:WIKI}, we ran SLNET-Metrics, \corpusSLNET{}'s metric collection tool, first on R2018b and then R2019b, as the latter ignores `resource' folder, which some older \corpusSLNET{} projects use. By following this process, we were able to reproduce their reported metrics.

Like \CorpusBoll{}, \corpusSLNET{} only offers project snapshots, but to assess its suitability for evolution studies, we require its git repositories. In April 2023 we obtained 208/225 SLNET GitHub repositories, as 17 projects were offline. We refer to this collection as \CorpusSLNETEvol{},  which we have made available for other researchers to analyze.

\subsection{Issues in \SL{} Tool-chain Found}

While trying to reproduce \corpusSLNET{}'s results, we encountered the following two \SL{} issues. \MW{} classified the first one as a bug and the second one as a documentation issue. First, when using multiple machines to speed up metric collection, \SL{}~R2018b crashed while compiling a \corpusSLNET{} model on Windows but compiled the model without issue on Ubuntu. We reported this issue (\#04254318), which \MW{} confirmed as a bug and fixed in \SL{} R2021b.

Finally, we reported (\#04386513) that the cyclomatic complexity definition of the multiport switch~\cite{doc:multiportswitch} did not seem to match \SL{}'s metrics results. MathWorks addressed this issue by updating its public metric description~\cite{doc:SL-McCabe}.

\section{Replicating Empirical Results Using \corpusSLNET{}}

To date, empirical data on \SL{} models and projects have been obtained on select closed-source projects and smaller open-source corpora (i.e., \CorpusCurated{} and \CorpusBoll{}). We would thus like to know how these earlier results generalize to the larger \corpusSLNET{} corpus of \SLNETlegalTotalProject{}~open-source projects and their~\SLNETlegalTotalModel{} \SL{} models. As earlier work has not characterized \corpusSLNET{}, we will first put it into context for any subsequent findings or comparisons.

As in similar comparative studies, when interpreting experimental results we need to know how much results are skewed by differences in experimental setups. While conceptually straight-forward, calculating \SL{} metrics is influenced by many parameters (Figure~\ref{fig:experiment-params}) and we realized that earlier studies did not document all relevant parameter values.

To increase confidence in our results we replicate earlier experiments where possible. Unless noted differently we apply the same metric extraction setup to all corpora---i.e., the same of our researchers use a single consistent set of metric definitions, metric tool version (SLNET-Metrics), \SL{} version (R2020b on Ubuntu~18.04), and toolboxes~\cite{wiki:replicationTool}. 

We used \SL{}~R2020b as it enhanced metric calculation~\cite{doc:enhancedcyclmaticcomplexity}. For example, in \SL{}~R2019b a video surveillance system's~\cite{repo:videosurvelliance} cyclomatic complexity is 38,403, which on manual inspection seems highly inflated. For the same system \SL{}~R2020b returns 322. Such a drastic change 
makes it hard to directly compare our results with results reported elsewhere, e.g., the \corpusSLNET{} work used \SL{}~R2019b~\cite{paper:slnet}.

\begin{insight}
Small changes in experimental setup can drastically skew \SL{} model metrics. In one example, upgrading to a newer version of \SL{} changed a model's cyclomatic complexity from 38,403 to 322.\\
\textbf{Implication:} There are subtle but severe pitfalls when comparing \SL{} metric results across papers. To increase confidence in such comparisons we thus repeat earlier experiments where possible.
\end{insight}

\subsection{Removing User-defined Libraries And Test Harnesses}

User-defined libraries and test harnesses serve different goals than regular \SL{} models. As they are also  structurally different, we first identify and separate them from the regular models. While user-defined libraries are interesting themselves, for analyzing regular models we treat user-defined libraries like all other libraries. We thus either inline blocks from all or none of the libraries. Following prior work~\cite{Chowdhury18Curated}, we use the \SL{} API~\cite{doc:bdIsLibrary} and identify 235 user-defined libraries in \CorpusCuratedRep{}, 411 in \CorpusBollRep{}, and 1,137 in \corpusSLNET{}.

\SL{}'s Test API~\cite{doc:sltest} can identify models as test harnesses and we thus remove 9 test harnesses from \corpusSLNET{} and two~each from \CorpusCuratedRep{} and \CorpusBollRep{}. This is likely an under-count, as many open-source projects may not have the license necessary for this API and thus use workarounds. We thus heuristically label (but not remove) models as potential test harnesses by checking if model and folder names contain ``test'' or ``harness'', thereby labeling 143~models in~\CorpusCuratedRep{}, 233~in~\CorpusBollRep{}, and 903~in~\corpusSLNET{}.

\subsection{RQ1: Basic \SL{} Model Metrics of Corpora}

At a high level, while it contains significantly more models,  \corpusSLNET{} is not a superset of the previous open-source corpora. Even when containing the same model, corpora may differ in the included model version, due to different corpus collection times. When treating all versions of a model as the same model and including user-defined library models, \corpusSLNET{} contains 
30\% of the \CorpusCurated{} models ($328/1071$), 
36\% for \CorpusCuratedRep{} ($402/1117$), 
28\% for \CorpusBoll{} ($492/1734$), and 
28\% for \CorpusBollRep{} ($492/1736$).

\begin{table*}[h!t]
    \centering
\caption{Model metrics after removing library \& test harness models in
\CorpusCuratedRep{} (top),
\CorpusBollRep{} (middle), and 
\corpusSLNET{} (bottom);
M~=~models;
Mc~=~models compiling in our setup; 
Mh~=~hierarchical models; 
C~=~non-hidden connections;
\textvisiblespace{}\textsuperscript{t0}~=~via \CorpusCurated{}'s metric tool;
var~=~variable;
nor~=~normal;
ext~=~external;
PIL~=~processor in the loop;
ac~=~accelerator;
rap~=~rapid accelerator;
Industry-M~=~Industry Mathworks;
M-Central~=~MATLAB Central;
excludes 14 \corpusSLNET{} models that crash \SL{}~R2020b;
includes 20~\corpusSLNET{} models for which \SL{}~R2020b does not show solver and simulation metrics.
  }  
      \resizebox{\textwidth}{!}{%
    \begin{tabular}{lrrrrrrrrrrrrrrr}
  \toprule
  \multirow{2}{*}{} 
      & 
      \multicolumn{2}{c}{Models} &
      \multicolumn{2}{c}{Hierarchical} &
      \multicolumn{2}{c}{Blocks} &
      \multicolumn{2}{c}{Connections} &
      \multicolumn{2}{c}{Solver Step} &  
      \multicolumn{5}{c}{Simulation Mode}  \\  
     &
     \multicolumn{1}{c}{M} & 
     \multicolumn{1}{c}{Mc} & 
     \multicolumn{1}{c}{Mh} & 
     \multicolumn{1}{c}{Mh\textsuperscript{t0}} & 
     \multicolumn{1}{c}{B} & 
     \multicolumn{1}{c}{B\textsuperscript{t0}} & 
     \multicolumn{1}{c}{C} & 
     \multicolumn{1}{c}{C\textsuperscript{t0}} &
      fixed & 
      \multicolumn{1}{c}{var} &
      \multicolumn{1}{c}{nor} & 
      ext & PIL & ac & rap\\
     \midrule
Tutorial &  41 & 41 & 37 & 40 & 3,703 & 13,917 & 3,700 & 14,020 & 13 & 28 & 41 & 0 & 0 & 0 & 0 \\
GitHub & 165 & 92 & 53 & 151 & 7,350 & 20,734 & 7,967 & 21,500 & 60 & 105 & 162 & 2 & 0 & 1 & 0\\
M-Central & 674 & 294 & 488 & 595 & 76,473 & 483,645 & 80,683 & 473,466 & 257 & 417 & 655 & 14 & 1 & 4  &0\\
SourceForge & 230 & 33 & 196 & 201 & 18,444 & 126,123 & 17,800 & 125,021 & 183 & 47 & 175 & 55 & 0 & 0 &0\\
Other & 7 & 4 & 3 & 7 & 611 & 680 & 636 & 701 & 1 & 6 & 7 & 0 & 0 & 0 & 0\\
        \midrule
$\sum$ \CorpusCuratedRep{} & 1,117 & 464 & 777 & 994 & 106,581 & 645,099 & 110,786 & 634,708 & 514 & 603 & 1,040 & 71 & 1 & 5   & 0\\                     
         \midrule
Academic & 690 & 232 & 456 & 634 & 75,813 & 185,574 & 86,223 & 185,733 & 229 & 461 & 597 & 68 & 0 & 16 & 9\\
Industry-M & 404 & 61 & 259 & 351 & 30,826 & 220,011 & 27,631 & 212,299 & 176 & 228 & 399 & 4 & 1 & 0 & 0\\
Industry & 174 & 15 & 93 & 161 & 24,753 & 180,929 & 25,116 & 194,655 & 135 & 39 & 169 & 3 & 0 & 1 & 1\\
No info & 55 & 24 & 44 & 46 & 4,889 & 26,690 & 5,524 & 26,803 & 29 & 26 & 54 & 1 & 0 & 0 & 0\\
\midrule
 $\sum$ \CorpusBollRep{} & 1,323 & 332 & 852 & 1,192 & 136,281 & 613,204 & 144,494 & 619,490 & 569 & 754 & 1,219 & 76 & 1 & 17 & 10\\
         \midrule         

GitHub & 1,637 & 541 & 875 & 1,297 & 190,213 & 424,175 & 188,069 & 400,753 & 860 & 759 & 1,498 & 103 & 2 & 14 & 2  \\     
 M-Central & 6,239 & 3,370 & 3,874 & 5,485 & 828,210 & 3,197,090 & 914,857 & 3,074,782 & 1,753 & 4,484 & 5,971 & 186 & 2 & 76   &   2  \\
        
        \midrule
$\sum$ \corpusSLNET{} & 7,876 & 3,911 & 4,749 & 6,782 & 1,018,423 & 3,621,265 & 1,102,926 & 3,475,535 & 2,613 & 5,243 & 7,469 & 289 & 4 & 90 & 4\\
         \bottomrule
    \end{tabular}
    }
\label{tab:replication}
\end{table*}

The remainder of this work removes from each corpus each model that is a test harness or a user-defined library. This differs from earlier work that treated user-defined library models as regular models and thus included them in overall metric counts~\cite{simulink:corpus:boll2021characteristics}. (The only exceptions are the three Table~\ref{tab:replication} \textvisiblespace{}\textsuperscript{t0} columns, which inline user-defined libraries.) Table~\ref{tab:replication} compares \CorpusCuratedRep{}, \CorpusBollRep{}, and \corpusSLNET{} on basic \SL{} model metrics, such as number of models, models that are hierarchical, blocks, connections, and solver and simulation modes.

\subsubsection{Model Size} 
A widely-used proxy for model size is the model's number of blocks~\cite{simulink:cyclo:Monika,simulink:test:testprioritization,simulink:empirical:SchlieWSCS17}.
For example, a recent paper conducted experiments on what it introduced as large industrial automotive models, containing 3.7k--73k blocks (and having hierarchy depth 8--16)~\cite{Pantelic18}. 
Boll \etal{} report conversations with \SL{} experts indicating typical industrial models often have 1k--10k blocks~\cite{simulink:onreplicability:Boll}. Industry-scale models at automotive supplier Delphi were earlier reported to have on average some 750~blocks~\cite{simulink:faultlocalization_Liu}.

Table~\ref{tab:replication} shows that (except for ``Others''), including imported library blocks (B\textsuperscript{t0}) at least doubles the overall block count.
Focusing on 1k+ block models, 
\CorpusCurated{}'s custom tool (which includes imported library blocks) found 93~such models in \CorpusCurated{} on Simulink R2017a. On Simulink 2020b, \CorpusCurated{}'s tool found 
132 such models in \CorpusCuratedRep{},
139 in \CorpusBollRep{}, and
799 in \corpusSLNET{}. 
When excluding any imported library blocks, 
\CorpusCuratedRep{} contains 14 such models,
\CorpusBollRep{}~15, and
\corpusSLNET{} 148.

\subsubsection{Hierarchical \& Compiling Models}

Model hierarchy is important for studying model complexity, model slicing and evaluating \SL{} model generation tools~\cite{lab:SLGPT,Chowdhury20SLEMI,SLEMI:Demo,simulink:slicing:ReichertG12}. \CorpusCuratedRep{} has 777 hierarchical models, of which we could compile 44\%. Of \CorpusBollRep{}'s 852 hierarchical models we could only compile 20\%. Of \corpusSLNET{}'s 4.7k hierarchical models we could compile 47\%. \CorpusBollRep{}'s low compile rate can be attributed to that corpus not distributing non-model files, which may have served as dependencies for the Simulink model.

\subsubsection{Project and Model Metric Distributions}

\begin{table*}[h!t]
    \centering
    \caption{Model (after removing library \& test harness models) metric distributions per project (p) and per model (m) in 
    \CorpusCuratedRep{} (R), 
    \CorpusBollRep{} (20R), and 
    \corpusSLNET{} (N);  
    Cyclom. C.~=~cyclomatic complexity (for a project the max of its models);
    Model Ref.~=~model references; 
    Alg. L.~=~algebraic loops;
    LL Blocks~=~library linked blocks;
    Sub. Blocks~=~blocks in a subsystem at depth that has most such blocks.
    }   
    \begin{tabular}{lc|rrr|rrr|rrr|rrr|rrr}
    \toprule
   \multicolumn{1}{l}{} &  
   & \multicolumn{3}{c|}{Min}
   & \multicolumn{3}{c|}{Max}
   & \multicolumn{3}{c|}{Average} 
   & \multicolumn{3}{c|}{Median}
   & \multicolumn{3}{c}{Standard Deviation} \\
    &   
    & \textsubscript{R} 
    & \textsubscript{20R} 
    & \textsubscript{N} 
    & \multicolumn{1}{c}{\textsubscript{R}}
    & \multicolumn{1}{c}{\textsubscript{20R}}
    & \multicolumn{1}{c|}{\textsubscript{N}}
    & \multicolumn{1}{c}{\textsubscript{R}}
    & \multicolumn{1}{c}{\textsubscript{20R}}
    & \multicolumn{1}{c|}{\textsubscript{N}}
    & \multicolumn{1}{c}{\textsubscript{R}}
    & \multicolumn{1}{c}{\textsubscript{20R}}
    & \multicolumn{1}{c|}{\textsubscript{N}}
    & \multicolumn{1}{c}{\textsubscript{R}}
    & \multicolumn{1}{c}{\textsubscript{20R}}
    & \multicolumn{1}{c}{\textsubscript{N}}
    \\
    \midrule

 Models & p & 1 & 1 & 1 & 124 & 124 & 237 & 5.6 & 6.9 & 2.8 & 1.0 & 1.0 & 1.0 & 14.7 & 16.4 & 9.7\\\midrule

\multirow{2}{*}{Blocks} 
& p & 1 & 1 & 0 & 13,555 & 13,831 & 172,196 & 457.4 & 706.1 & 362.8 & 116.0 & 140.0 & 52.0 & 1,419.9 & 1,959.8 & 3,577.1\\
& m & 1 & 0 & 0 & 13,555 & 13,555 & 18,255 & 95.4 & 103.0 & 129.3 & 25.0 & 25.0 & 27.0 & 448.4 & 430.6 & 690.1\\\midrule

\multirow{2}{*}{Block types} 
& p & 1 & 1 & 1 & 55 & 58 & 104 & 18.3 & 19.2 & 13.2 & 16.0 & 17.0 & 11.0 & 11.1 & 11.9 & 9.3\\
& m & 1 & 1 & 1 & 47 & 47 & 101 & 10.6 & 10.4 & 10.4 & 8.0 & 8.0 & 8.0 & 8.3 & 7.8 & 7.9\\\midrule

\multirow{2}{*}{Connections} 
& p & 0 & 0 & 0 & 14,169 & 16,491 & 231,672 & 475.5 & 748.7 & 392.9 & 124.0 & 153.0 & 57.0 & 1,422.1 & 2,103.5 & 4,611.7\\
& m & 0 & 0 & 0 & 14,169 & 14,169 & 25,078 & 99.2 & 109.2 & 140.0 & 26.0 & 27.0 & 28.0 & 466.2 & 453.7 & 887.1\\\midrule

\multirow{2}{*}{Subsystems} 
& p & 0 & 0 & 0 & 1,809 & 1,873 & 19,622 & 46.9 & 68.4 & 34.0 & 7.0 & 7.0 & 2.0 & 179.3 & 210.8 & 414.1\\
& m & 0 & 0 & 0 & 1,294 & 1,294 & 2,117 & 9.8 & 10.0 & 12.1 & 3.0 & 2.0 & 2.0 & 44.1 & 41.8 & 75.3\\\midrule

\multirow{2}{*}{Cyclom. C.} 
& p & 0 & 0 & 0 & 322 & 322 & 2,404 & 27.7 & 30.7 & 22.2 & 7.0 & 7.0 & 5.0 & 49.4 & 54.1 & 81.1\\
& m & 0 & 0 & 0 & 322 & 322 & 2,404 & 14.0 & 13.6 & 13.7 & 4.0 & 4.5 & 2.0 & 32.4 & 31.6 & 59.0\\\midrule

\multirow{2}{*}{Model Ref.} 
& p & 0 & 0 & 0 & 4 & 10 & 54 & 0.1 & 0.1 & 0.1 & 0.0 & 0.0 & 0.0 & 0.4 & 0.8 & 1.5\\
& m & 0 & 0 & 0 & 4 & 2 & 12 & 0.0 & 0.0 & 0.0 & 0.0 & 0.0 & 0.0 & 0.2 & 0.1 & 0.4\\\midrule 

\multirow{2}{*}{Alg. L.} 
& p & 0 & 0 & 0 & 7 & 9 & 37 & 0.2 & 0.2 & 0.1 & 0.0 & 0.0 & 0.0 & 0.7 & 1.0 & 1.1\\
& m & 0 & 0 & 0 & 2 & 1 & 6 & 0.1 & 0.1 & 0.1 & 0.0 & 0.0 & 0.0 & 0.2 & 0.2 & 0.3\\\midrule

\multirow{2}{*}{LL Blocks} 
& p & 0 & 0 & 0 & 657 & 423 & 2,311 & 9.1 & 11.8 & 5.6 & 0.0 & 0.0 & 0.0 & 53.8 & 48.7 & 81.1\\
& m & 0 & 0 & 0 & 31 & 31 & 441 & 1.9 & 1.7 & 2.0 & 0.0 & 0.0 & 0.0 & 4.5 & 4.3 & 15.0\\\midrule

Sub. Blocks & - & 2 & 2 & 3 & 21 & 21 & 100 & 9.6 & 9.5 & 9.1 & 11.0 & 11.0 & 7.0 & 3.9 & 3.9 & 11.5\\

\bottomrule
    \end{tabular}
    \label{tab:metric_distribution}
\end{table*}

Table~\ref{tab:metric_distribution} shows model metric distributions across \CorpusCuratedRep{}, \CorpusBollRep{}, and \corpusSLNET{}. 
The majority of \corpusSLNET{} models are relatively small, with mean exceeding median values. The overall distribution of metrics in \corpusSLNET{} is akin to that of earlier corpora, i.e., offering a broad spectrum with most standard deviations exceeding the means. \corpusSLNET{} however offers a broader range of \SL{} models with similar min but notably larger max metric values. Following are additional distribution details of project size, most frequently used block types, and file types.

\paragraph{Project size}
Similar to earlier corpora, the distribution of models in \corpusSLNET{} is skewed towards a few large projects. The 50 largest projects (i.e., the largest 1.8\% of projects) contain 35\% of all models, while 76\% of the projects contain just one model. Some \corpusSLNET{} projects feature 18 empty models alongside non-empty models. By comparison, in \CorpusBollRep{}, 5/194 projects contain 35\% of the models, and 53\% of the projects contain just one model. With the exception of a single \corpusSLNET{} project that comprises a library model, all projects include some blocks and signal lines. 

\begin{figure}[h!]
\centering
\begin{subfigure}[b]{\linewidth}
\centering
\includegraphics[width=0.85\linewidth]{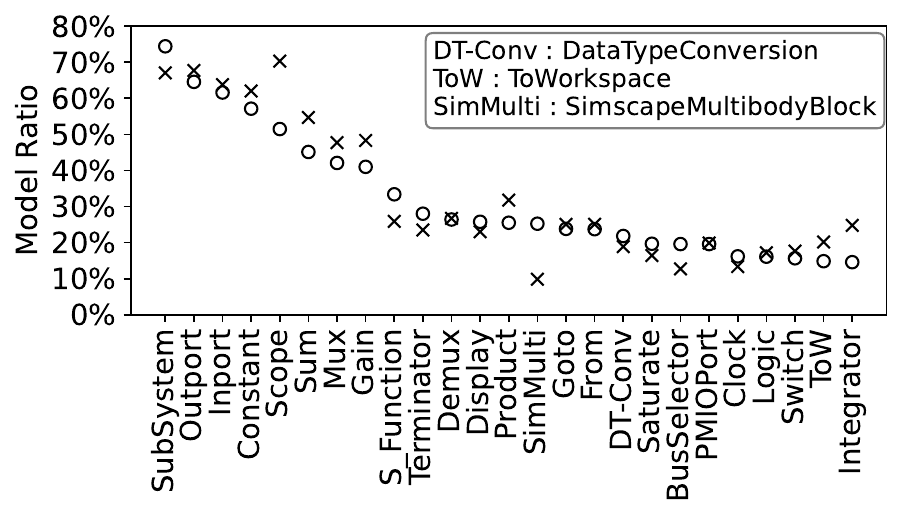}
\caption{Most-common block types in \CorpusCuratedRep{} (o) and their \corpusSLNET{} (x) rate.       
}
\label{fig:freq_blk_orderedByslc2}
\end{subfigure}

\begin{subfigure}[b]{\linewidth}
\centering
 \includegraphics[width=0.85\linewidth]{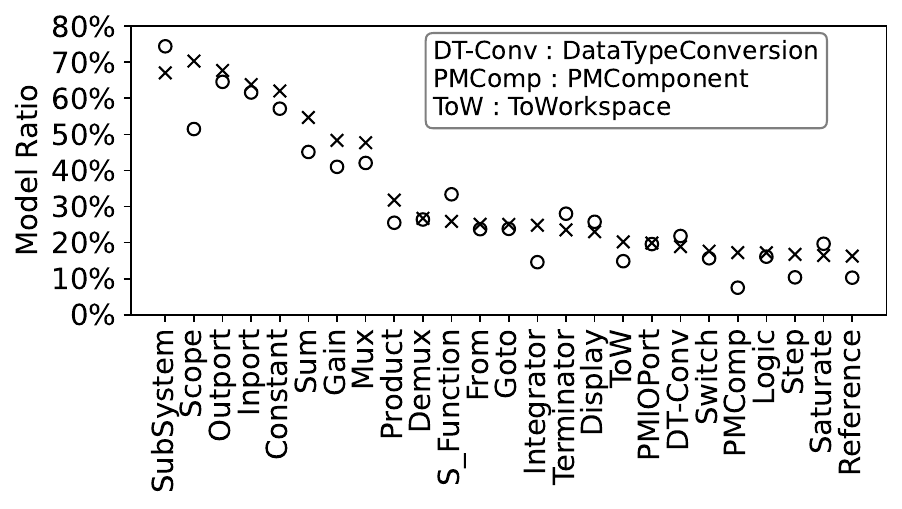}
\caption{Most-common block types in \corpusSLNET{} (x) and their \CorpusCuratedRep{} (o) rate. }
\label{fig:freq_blk_orderedByslnet}
\end{subfigure}
\caption{
Most-common block types in \CorpusCuratedRep{} (a) and \corpusSLNET{} (b).    
}
 \label{fig:freq_blk}
\end{figure}

\paragraph{Most Frequently Used Block Types}
  
Figures~\ref{fig:freq_blk_orderedByslc2} and~\ref{fig:freq_blk_orderedByslnet} show that the distributions of the most-commonly used block types are similar in \CorpusCuratedRep{} and \corpusSLNET{}. For example, in each corpus over 60\% of models contain a SubSystem block, making SubSystem appear in the most models in both corpora. \corpusSLNET{} uses SubSystem less-widely, likely as 28\%~of \corpusSLNET{} models have less than 8 blocks, which typically does not require a SubSystem block.

\CorpusCuratedRep{} models use 156 distinct block types vs. 203 in \corpusSLNET{} (150 are in both). \corpusSLNET{} thus offers a potentially valuable resource for research studies~\cite{simulink:onthefly_sanchez,chowdhury2018icse}. Both \CorpusCurated{} and \CorpusBoll{} studies included library-imported blocks and reported a lower occurrence of output blocks (e.g., Scope~\cite{doc:scope}, Display~\cite{doc:display}, and ToWorkspace~\cite{doc:toworkspace}) than \corpusSLNET{}. The possible explanation for this discrepancy is that, like in procedural programming languages (where programmers include logging statements at various execution points), libraries may not have such statements for efficiency purposes. This practice is also observed in \SL{} modeling.

Furthermore, From~\cite{doc:from} and Goto~\cite{doc:goto} blocks, which are typically used to improve the visual layout of the model, are equally widely used in \CorpusCurated{} and \corpusSLNET{}. 
However, excessive non-local usage of From and Goto blocks adversely affects readability and design, warranting further investigation.

\paragraph{File types} 

Each \SL{} model is stored in one of two file formats, the MDL legacy file format or SLX. Introduced in \SL{} R2012a, SLX conforms to the Open Packaging Conventions (OPC) interoperability standard. Across corpora, few projects contain both MDL and SLX files (\CorpusCuratedRep{} 3\%, \CorpusBollRep{} 7\%, and \corpusSLNET{} 2\%). Overall the major file type has shifted from MDL in \CorpusCuratedRep{} to SLX in \corpusSLNET{} (39\% of \CorpusCuratedRep{} models are in SLX, 45\% for \CorpusBollRep{}, and 55\% for \corpusSLNET{}). The prevalence of SLX files in open-source models is significant for developing SLX to MDL back-transformation tools~\cite{AdhikariRS21BackwardVersion}.

In summary, \corpusSLNET{} shares many similarities with prior corpora and offers a broader view of open-source \SL{} projects. 
The majority of \corpusSLNET{} models are small, which may be relevant for analyzing simple models~\cite{lab:SLGPT,Shrestha19SRC,lab:DeepFuzzSL,simulink:tooltoimproveSimulinkPractices}, while also including a substantial number of non-trivial models using diverse features.

\begin{insight}
 As in many other kinds of open-source projects~\cite{github:perilsofmininggithub,MaCZX16DominantGitHubProjects}, \corpusSLNET{} project and model metrics follow long-tailed distributions. \\
 \textbf{Implication:} 
Research studies may use \corpusSLNET{} subsets based on their objectives. The diverse \corpusSLNET{} corpus can help address generalizability challenges in model-based development research.
\end{insight}

\section{Replicating Findings on Modeling Practices}
\label{sec:slnetconfirnandcontradicts}

\subsection{Converging Result: Model Referencing} 

Analogous to classes in object-oriented programming, model references~\cite{doc:mdlref} enable modular model design, unit testing, and code reuse. But similar to the \CorpusCurated{} work~\cite{Chowdhury18Curated}, we found that only 
10~\CorpusCuratedRep{} (0.9\%),
18~\CorpusBollRep{} (1.4\%), and
139~\corpusSLNET{} models (1.8\%) use model referencing. Even when accounting for the skewed \corpusSLNET{} model size distribution, Table~\ref{tab:metric_distribution} shows that model reference use remains sparse.

\subsection{Converging Result: Algebraic Loops}
An algebraic loop arises from a circular dependency between a block's output and input at the same simulation time step. An algebraic loop may reduce simulation performance or prevent the solver from resolving the loop. As the \CorpusCurated{} work~\cite{Chowdhury18Curated}, we found such loops relatively rarely, with only 20~\CorpusCuratedRep{} and 186~\corpusSLNET{} models containing such loops.

\subsection{Converging Result: Small Class Phenomenon} 

Zhang \etal{} observed the ``small class'' phenomenon in Java programs (most classes have few lines of code while a few classes are large) and found a high correlation between class size and number of defects~\cite{java:empirical:zhang,java:zhang2009distribution}. In \SL{}, subsystems are used to encapsulate a function, resulting in a hierarchical model. Similar to the small class phenomenon noted in the \CorpusCurated{} work~\cite{Chowdhury18Curated}, we observe that the median number of blocks in a subsystem at any hierarchy does not exceed 11 in both \CorpusCuratedRep{} and \corpusSLNET{}. This may inform future hypotheses on \SL{} subsystem size and defects.

\begin{insight}
    The median number of blocks in a subsystem at any hierarchy level does not exceed 11.\\ 
    \textbf{Implication:} More research is needed to assess how subsystem size impacts \SL{} model quality. 
\end{insight}

\subsection{Converging Result: S-function Reuse Rate}

\begin{table}[h!]
    \centering
    \caption{S-function per-model reuse rate for models with 1+ S-functions; 
    M\textsubscript{S-fct}~=~models with 1+ S-functions;
    LQ~=~lower quartile;
    UQ~=~upper quartile;
    med~=~median.
    }
    \begin{tabular}{l|r|rrrrrr}
    \toprule
         & M\textsubscript{S-fct} & min & LQ & med & UQ & max & avg\\
       \midrule
       \CorpusCuratedRep{} & 351 & 0.0 & 0.0 & 0.0 & 0.38 & 0.92 &0.20\\
       \CorpusBollRep{} & 378 &0.0 & 0.0 & 0.0 & 0.50 & 0.98 & 0.23\\
       \corpusSLNET{} & 1,504 & 0.0 & 0.0 & 0.0 & 0.50 & 0.99 & 0.21\\
         \bottomrule
    \end{tabular}
    \label{tab:sfunc_reuse_rate}
\end{table}
Besides reuse of legacy C code, S-functions allow within-model code reuse (i.e., defined once but added to and used in several model components). In the same spirit as the \CorpusCurated{} work~\cite{Chowdhury18Curated}, Table~\ref{tab:sfunc_reuse_rate} shows that S-functions are not widely used, with just 31\% of \CorpusCuratedRep{} models and 20\% of \corpusSLNET{} using S-functions. For models that use S-functions, 41\% of \CorpusCuratedRep{} models and 40\% of \corpusSLNET{} models reuse at least one S-function (but these models' median S-function reuse rate is zero across corpora).


\subsection{Diverging Result: Cyclomatic Complexity vs Other Metrics}

We conduct a correlation analysis between cyclomatic complexity and the other Table~\ref{tab:CycloCorrelation} model metrics using Kendall's $\tau$. We only use models for which we could calculate cyclomatic complexity (e.g., excluding models we could not compile). As in the \CorpusCurated{} study, for \CorpusCuratedRep{} we used non-Simple models. For \CorpusBollRep{}, we used industry and industry-MathWorks models. As \corpusSLNET{} models are not categorized, we used those containing 200+ blocks. All metrics exhibit a statistically significant correlation at a 0.05 significance level.

\begin{table}[h]
\centering
\caption{Correlation between cyclomatic complexity and model metrics;
M,B,C from Table~\ref{tab:replication}: models, blocks, and non-hidden connections;
UB~=~unique block types;
MHD~=~max. hierarchy depth;
CRB~=~child-model representing blocks i.e., model reference and
subsystem;
NCS~=~contained subsystems.
}  
\begin{tabular}{lr|rccccc}
\toprule
& \multicolumn{1}{c|}{M} & \multicolumn{1}{c}{B} & C & UB  & MHD & CRB &  NCS \\
\midrule
 \CorpusCuratedRep{}   &   160 & 0.29& 0.32          & 0.31 &   \textbf{0.38} & 0.28 & 0.29\\
\CorpusBollRep{} & 58 & 0.16 & 0.16 & 0.20 & 0.31 &\textbf{0.41} &\textbf{0.41} \\    
\corpusSLNET{}\textsubscript{$\ge$200} & 279 & 0.27 & 0.27 &  0.23 & 0.10 & \textbf{0.28} & 0.27 \\   
 \corpusSLNET{}\textsubscript{200-300} & 111 & -0.02 & 0.12 &  \textbf{0.16} &  0.05 & 0.07 & 0.07 \\
\bottomrule
\end{tabular}
\label{tab:CycloCorrelation}
\end{table} 

\CorpusCuratedRep{} models have a weak positive correlation (0.28 to 0.38) between cyclomatic complexity and model metrics. For \CorpusBollRep{} models the correlation is positive and weak to (barely) moderate (0.16 to 0.41).
For \corpusSLNET{} models with 200+ blocks the correlation is positive but remains weak (0.10 to 0.28).  

\begin{insight}
Contrary to previous work~\cite{simulink:Marta}, cyclomatic complexity does not seem strongly
correlated with other model metrics.\\
\textbf{Implication:} Similar to Java- and C-like languages, in \SL{} cyclomatic complexity seems to remain an independently valuable metric.
\end{insight}

\begin{table*}[h!t]
\centering
    \caption{\CorpusSLNETEvol{} and \CorpusBollEvol{} per-model (m) and per-project (p) change metrics;
    Total commits, 
    commits per day during project duration,
    merge commits ($\succ$), and
    commits of 1+~mdl/slx files (MS);
    commit authors and
    commit\textsubscript{MS} authors;
    med~=~median;
    std~=~standard deviation.}
    \begin{tabular}{lcrrrrr|rrrrr}
\toprule
& & \multicolumn{5}{c}{\CorpusBollEvol{}} & \multicolumn{5}{|c}{\CorpusSLNETEvol{}}\\
 & 
 & \multicolumn{1}{c}{min}
 & \multicolumn{1}{c}{max}
 & \multicolumn{1}{c}{avg}
 & \multicolumn{1}{c}{med}
 & \multicolumn{1}{c}{std} 
  & \multicolumn{1}{|c}{min}
 & \multicolumn{1}{c}{max}
 & \multicolumn{1}{c}{avg}
 & \multicolumn{1}{c}{med}
 & \multicolumn{1}{c}{std} \\
\midrule                                                                
Commits & p                                 &1	     &590	&62.7	&10.5	&124.6           &1	   &963	       &43.9	    &7.5	&120.1\\
Commit / day & p                            &0	     &4	     &0.9	&0.3	&1.2             &0	   &24	       &1.9	        &0.6	&3.1\\
Commits\textsubscript{MS} [\%] & p          &0	     &100    &38.8	&26.8	&31.2            &1	   &100	   &31.4	    &25.0	&23.5\\
Commits\textsubscript{$\succ$} [\%] & p     &0	     &17	&2.7	&0.0	&4.7            &0	   &40	   &3.2	        &0.0	&6.7\\
Updates\textsubscript{MS} & m                &0	     &43	&3.3	&1.0   	&5.7             &0	   &53	       &1.8	        &1.0	    &2.8\\
\multirow{2}{*}{Authors} &p                 &1	     &16    &2.8	&2.0	&3.5             &1	   &21	       &2.0	        &1.0	&2.6\\
         & m                                &1	    &3	   &1.1	   &1.0	   &0.4             &1	   &8	       &1.3	        &1.0	    &0.7\\
Authors\textsubscript{MS}  [\%] & p         &0	     &100	&68.6	&75.0	&34.5            &10	&100	   &82.2	    &100.0	&26.1\\
\bottomrule
    \end{tabular}
    \label{tab:proj_evol}
\end{table*}

\subsection{Converging Result: Suitability For Change Studies}
\label{sec:SimulinkProjectEvolution}

To assess their applicability for \SL{} model and project change studies, we analyzed \CorpusBollEvol{}'s 32 and \CorpusSLNETEvol{}'s 208 git repositories (for \CorpusSLNETEvol{} we only studied the commits until \corpusSLNET{}'s February 2020 snapshot). Three projects (with 811 commits) were in both corpora. 

Table~\ref{tab:proj_evol} gives an overview of the project and model change metrics. For example, 53\% of \CorpusBollEvol{} projects (17/32) and 39\% of \corpusSLNET{} projects (82/208) are maintained by at least two collaborators, of which 8/17 and 32/82 have commits spanning over a year. Just 22\% of \CorpusBollEvol{} and 15\% of \CorpusSLNETEvol{} projects have more than 50 commits. Across \CorpusBollEvol{} and \CorpusSLNETEvol{} projects, 20\% of commits involved updates or the creation of one or more models.

In both corpora, an average of 22\% of models were under active development throughout the projects with 3+ commits, indicating the models were primary artifacts of these projects. However, 40\% of \CorpusBollEvol{} and almost half of \CorpusSLNETEvol{} projects did not update their models after committing them to the repository. In both corpora, roughly 55\% of models were not updated at all. The lack of model updates may be due to \GH{} \SL{} projects mainly serving as archives---like most other \GH{} projects~\cite{github:perilsofmininggithub}.

\begin{figure}[h!t]
    \centering
    \begin{subfigure}[b]{.9\linewidth}
         \includegraphics[trim = .3in 0in .6in .3in, clip,width=\linewidth]{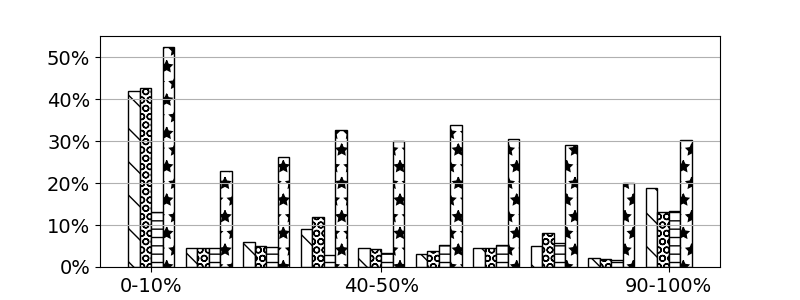}
        \caption{Timeline of 26/32 \CorpusBollEvol{} 3+ commit projects.}
        \label{fig:bollEvol}
    \end{subfigure}
    \begin{subfigure}[b]{.9\linewidth}
    \includegraphics[trim = .3in 0.1in .6in .3in, clip,    width=\linewidth]{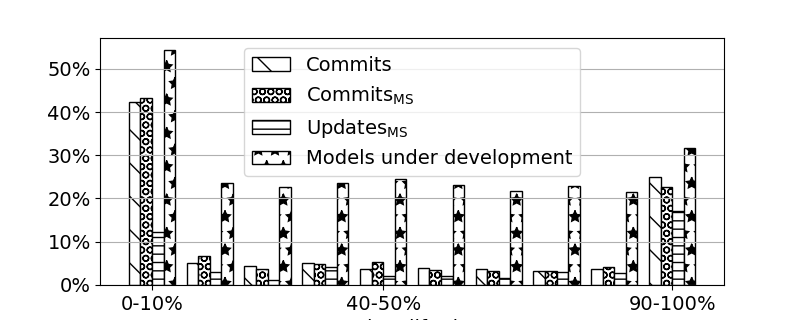}
      \caption{Timeline of 186/208 \CorpusSLNETEvol{} 3+ commit projects.}
        \label{fig:slnetEvol}        
    \end{subfigure}
\caption{
Across normalized project duration (x-axis): 
Total project commits,
commits of 1+ mdl/slx files,
individual mdl/slx file updates, and
mdl/slx files under development (i.e., in between a file's first and last commit).
}
\label{fig:project_evolution}
\end{figure}

Figure~\ref{fig:project_evolution} breaks each project’s duration into 10 buckets of equal length (normalized to each project’s duration). Here project
duration is the duration from a project’s first to last commit as recorded by the timestamps assigned by the authors’ machines. While this approach has its pitfalls, the more-active projects are usually less affected and we performed the basic recommended sanity checks to ensure there are no impossible outliers (e.g., commits with Unix time zero)~\cite{Flint22Pitfalls}.

To avoid potential skewing caused by ``code dump'' projects, Figure~\ref{fig:project_evolution} excludes projects with less than 3 commits, yielding 26 \CorpusBollEvol{} projects and 186 \CorpusSLNETEvol{} projects. Even with this filtering, the figure may still be biased towards projects with fewer commits as the majority of both \CorpusBollEvol{} and \CorpusSLNETEvol{} projects have less than 11 commits.

\begin{insight}
  A quarter of \CorpusSLNETEvol{} projects are developed collaboratively and have 1+ multi-revision models.
  \\
    \textbf{Implication:} \CorpusSLNETEvol{} projects have the potential to yield valuable insight into open-source \SL{} development.
\end{insight}

\subsection{Diverging Result: Open-source Code Generation Models} 
\label{sec:codegeneratemodel}

\SL{} models that can generate code are of interest in model-based research and tool-development~\cite{simulink:test:testprioritization,simulink:codegeneration:krizan,simulink:testsuitegen:Matinnejad,simulink:codegeneration:Mozumdar,simulink:codegeneration:cocosim}. Initially we applied \CorpusBoll{}'s heuristics to search for Embedded Coder~\cite{doc:embedded_coder} or TargetLink~\cite{doc:targetlink} traces. But we found inconsistencies between \CorpusBoll{}'s results (finding no code generation models) and their replication package's heuristics~\cite{dataset:boll-v1}. During our interactions the \CorpusBoll{} team acknowledged a bug and fixed it in their replication package version 2~\cite{dataset:boll}.

\begin{table}[]
    \centering
      \caption{Models configured for code generation;
      M~=~all models;
      EC\textsubscript{20}~=~\CorpusBoll{} Embedded Coder heuristics;
      EC~=~our Embedded Coder heuristics;
      GRT~=~Simulink Coder (Real-Time Workshop);      
      Other~=~other code generation toolboxes.
}
    \begin{tabular}{lr|r|rrr|r}
    \toprule
    & \multicolumn{1}{c|}{M}
    & EC\textsubscript{20} 
    & EC 
    & GRT 
    & Other 
    & Total  \\
    \midrule
    Tutorial &  41 & 1 & 1 & 12 & 0 & 13\\
    GitHub & 165 & 0 & 4 & 52 & 4 & 60\\
    MATC &674 & 0 & 47 & 101 & 109 & 257\\
    Sourceforge & 230 & 0 & 0 & 96 & 87 & 183\\
    Others &  7 & 0 & 0 & 1 & 0 & 1\\
    \midrule
    $\sum$ \CorpusCuratedRep{} & 1,117 & 1& 52 & 262  & 200 & 514\\
\midrule
    Academic &690 & 0 & 3 & 94 & 136 & 233\\
Industry-M & 404 & 0 & 33 & 77 & 67 & 177\\
Industry & 174 & 0 & 5 & 129 & 1 & 135\\
No Info & 55 & 0 & 1 & 28 & 0 & 29\\
    \midrule
    $\sum$ \CorpusBollRep{} & 1,323 & 0 & 42 & 328 & 204 & 574 \\
\midrule
    GitHub & 1,637 & 14 & 129 & 502 & 234 & 865\\
    MATC & 6,239 & 19 & 423 & 1,050 & 297 & 1,770\\
    \midrule
    $\sum$ \corpusSLNET{} & 7,876 & 33 & 552 & 1,552 & 531 & 2,635 \\
\bottomrule
\end{tabular}
\label{tab:codegeneratetraces}
\end{table}

Specifically, \CorpusBoll{}'s heuristics determine if a model can generate code based on the presence of \textit{atomic} subsystems~\cite{doc:atomicSubsystem} or special TargetLink blocks. This found 33~\corpusSLNET{} models configured for Embedded Coder but no TargetLink traces. 
We found this heuristic restrictive and not specific to Embedded Coder. Our counter-example model had non-atomic subsystems and successfully generated code via Embedded Coder. 

For background, while every \SL{} model can generate code using \SL{} Coder~\cite{doc:simulinkcoder}, this requires a fixed-step solver, which conflicts with the default variable-step solver model configuration. \SL{} models further rely on a target language compiler (TLC) file~\cite{doc:targetompilerbasiccs} to map \SL{} blocks and parameters to the target language's constructs.

\SL{} offers a set of standard-named TLC files that support various solver types~\cite{doc:selectatarget}. For example,`rsim.tlc' supports fixed-step and variable-step solvers. To determine if the \SL{} model is configured to generate code,  we follow a heuristic approach. First, we check if the model's TLC file name matches with one provided by \SL{} and the model is configured with appropriate solver type. Second, in cases where the solver type required is ambiguous, we make a conservative assumption that the model must be configured with the fixed-step solver. 

Table~\ref{tab:codegeneratetraces} shows the number of models configured for code generation. \corpusSLNET{} has 2,635 models with code generation capabilities, at least 4$\times$ more than previous corpora.


\begin{insight}
\corpusSLNET{} has 4$\times$ models configured for code generation (a common configuration in industrial models) than the largest earlier open-source model collection. 
\textbf{Implication:} Additional investigation is required to determine if the code generation models in \corpusSLNET{} can meet requirements of research studies. 
\end{insight}

\section{Threats to Validity}

Internal validity concerns the experimental design, data collection and analysis. In our replication efforts, we closely adhered to the original study's setup and tools. We calibrated the provided tools and contacted the authors for clarification  and consistency in data analysis. It is important to note that the choice of Simulink version can impact model metrics and introduce slight differences in insights.

Specifically, for a subset of 554~\corpusSLNET{} models (the models of the 10~\corpusSLNET{} projects with the most models) we compared model metrics obtained using both R2020b and R2022b. Results for all metrics were the same for all models, except for 3/554 models where the cyclomatic complexity differed by 2--6 between R2020b and R2022b.

External validity examines the generalizability of reproduced and replicated study results. In our case, the generalizability of our findings is limited to Simulink models within the \corpusSLNET{} corpus. \corpusSLNET{} may not represent all available Simulink projects, as its construction involved a keyword search on GitHub and filtering for redistributable projects. However, considering that the majority of results from the original studies, which involved some level of cherry picking in their corpus, hold true in \corpusSLNET{}--a larger dataset encompassing diverse models with a small overlap--we are optimistic in the generalizability of the presented results to other open-source Simulink models.

Construct validity ensures that the measures and metrics used in the replicated study accurately capture the intended concepts. Our confirmatory replication study inherits limitations from the original studies, such as not analyzing  Stateflow blocks or MATLAB code, which can contribute to the project's complexity. Also, \CorpusCurated{}'s heuristic used to identify test harnesses may have limitations, as manual inspection revealed 10\% of such models are test harnesses. Upon noticing issues with \CorpusBoll{}'s code generation heuristic, we proposed new methods after consulting with the original authors.

Reliability refers to the replicability of a study for obtaining same or similar results. To mitigate reliability risks, we distribute our analysis tool and complete replication package as open-source via permanent storage locations~\cite{dataset:simulinkanalysisdata, doc:slreplicationTool}. We encourage replication of our findings.

\section{Conclusions 
and Future Work}

The study investigated the reproducibility of previous empirical studies of \SL{} models and evaluated the generalizability of their results to the larger \corpusSLNET{} corpus. The \corpusSLNET{} study confirmed and contradicted earlier findings, highlighting its potential as a valuable corpus for model-based development research and also provided actionable insights for future research. We found that open-source \SL{} models generally follow good modeling practices and that few open-source models are comparable in size and properties to proprietary models. To that end, we proposed a heuristic to determine code generating \SL{} models. We also provided 208 Git repositories to facilitate model evolution studies. 

While this paper only analyzes Simulink model metrics focusing on reproducibility and replication, future work includes examining if the model metrics can be used to make predictions of process 
metrics such as defect prediction.

\section*{Acknowledgements}
Christoph Csallner has a potential research conflict of interest due to a financial interest with Microsoft and The Trade Desk. A management plan has been created to preserve objectivity in research in accordance with UTA policy. This material is based upon work supported by the National Science Foundation (NSF) under Grant No. 1911017 and a gift from MathWorks.

\bibliographystyle{IEEEtranS}
\bibliography{IEEEfull,ref}

\end{document}